\documentclass[runningheads,a4paper]{llncs}

\usepackage{amssymb}
\usepackage{graphicx}

\usepackage{url}
\urldef{\mailsa}\path|eya.benahmed@gmail.com|
\urldef{\mailsb}\path|ms.gouider@isg.rnu.tn|
\newcommand{\keywords}[1]{\par\addvspace\baselineskip
\noindent\keywordname\enspace\ignorespaces#1}
\newtheorem{Example}{Example}

\begin{document}

\mainmatter  % start of an individual contribution

\title{Towards an incremental maintenance of cyclic association rules}

\titlerunning{Towards an incremental maintenance of cyclic association rules}

\author{Eya BEN AHMED%
%\thanks{Please note that the LNCS Editorial assumes that all authors have used
%the western naming convention, with given names preceding surnames. This determines
%the structure of the names in the running heads and the author index.}%
\and
Mohamed Salah GOUIDDER
}
\authorrunning{Towards an incremental maintenance of cyclic association rules }

\institute{Higher Institute of Management of Tunis, TUNISIA\\
\mailsa\\
\mailsb\\
%\url{http://www.springer.com/lncs}
}


\toctitle{Towards an incremental maintenance of cyclic association rules}
\tocauthor{Authors' Instructions}
\maketitle

\begin{abstract}

Recently, the cyclic association rules have been introduced in order to discover rules from items characterized by their regular variation over time. In real life situations, temporal databases are often appended or updated. Rescanning the whole database every time is highly expensive while existing incremental mining techniques can efficiently solve such a problem. In this paper, we propose an incremental algorithm for cyclic association rules maintenance. The carried out experiments of our proposal stress on its efficiency and performance.
\end{abstract}

%\vspace{-0.6cm}

\section{\textbf{Introduction}}
%\vspace{-0.3cm}

\par Knowledge management is the most time consuming and expensive part of our daily lives \cite{ARM09}.
In fact, it is crucial to explore a valid knowledge at any moment \cite{Swami}. Mining association rules can achieve this overarching goal
however this task becomes intensely more complicated in front of the wide size of databases, calling up gigabyte, terabyte, or even larger, in some applications \cite{Sri94}. In this respect, extracting association rules has outstandingly grasped the interest of the data mining community. Through time, the volume of information increases, the databases must be updated with the new amounts of data \cite{Lee98}.
Considering that an association rule generates explicitly reliable knowledge according to an explored database at an accurate time. So that, each update of the database radically overwhelms the already stored patterns. A projection of the database's changes must be drawn on extracted association rules \cite{ARM09}.
Since then, several proposals to solve this problem appeared \cite{Lee98,Che96,26,Zha03,Lee97maintenanceof}.

Parallel to those efforts, cyclic mining of association rules was also investigated on several studies. Such investigations can be found in \cite{CAR98,EYA10a,EYA10b}.
The problem of cyclic association rules mining consists in generation of association rules from  articles  characterized by  regular  cyclic  variation  over
time. In \cite{CAR98}, Ozdon et al. presented the first strategies of cyclic association rules extraction. Then, as a response to the anomalies characterizing the already proposed approaches, a more efficient algorithm was introduced by Ben Ahmed et al. \cite{EYA10a}.
\par It can be seen that the research community has proposed separate solutions for the incremental mining and the cyclic association rules mining problems.
%However, incremental mining of cyclic association rules is motivated by the fact that soon, cyclic association rules must be maintained according to database updates.
\par In this paper, we present a new algorithm called \textsc{IUPCAR} (Incremental UPdate of Cyclic Association Rules) for incremental mining of cyclic association rules. This algorithm provides the benefits of fast incremental mining and efficient cyclic association rules extraction.
\par The rest of the paper is organized as follows: section 2 studies the fundamental bases on which our proposal is built.
Section 3 presents a formal description of the problem. Section 4 details our proposal and offers an illustrative example to scrutinize the mechanism of our approach. Carried out experiments stressing on the efficiency of our proposal are sketched in section 5. Finally, section 6 concludes the paper and points out avenues for future work.
%\vspace{-0.7cm}
\section{\textbf{Foundations of the proposed algorithm}}
%\vspace{-0.5cm}
\par This section presents the previous theoretical foundations which form the bases on which our proposed algorithm is built. These bases include the theoretical foundations of the two problems of cyclic association rules mining and incremental mining of association rules. Therefore, we briefly discuss the cyclic association rules. Finally, we describe the incremental mining problem.
%\vspace{-0.7cm}

\subsection{\textbf{Cyclic association rules}}
%\vspace{-0.32cm}
\par Considering temporal transactional databases, several temporal patterns can be extracted \cite{Vin07}. Typically, one of them is related to cyclic association rules.
The main idea behind this latter is to extract correlations between products which vary in a cyclic way. Thanks to its comprehension of the data, we presume that the user is the most privileged to fix the best length of cycle according to considered products.
Hence we can analyze on depending on monthly, weekly, daily or even hourly sales of products according to the length of cycle.

For example, let \emph{\textbf{September and October sales}} in a bookstore transaction database be shown in Table \ref{bookstore}.
In fact, we assume at the beginning of September according to the transactions \textbf{\emph{9.1}} and \textbf{\emph{9.2}}, that the sales of \emph{books} and \emph{notebooks} are highly important. This can be explained by the start of the academic year. Besides, this correlation is underlined also at the beginning of February by transactions \textbf{\emph{2.1}} and \textbf{\emph{2.2}}. This fact is due to the start of the second semester of the academic year. Effectively, the considered cycle in this case is the length of the semester namely six months.
The outcomes of these analysis are then used to support various business decisions in this bookstore.
 %\vspace{-0.6cm}
\par \begin{table}[htpd]
\begin{center}
\begin{tabular}{| c | c | c |}
\hline \textbf{Transaction ID} & \textbf{Items}\\
\hline \hline 9.1 & \emph{books} , \emph{notebooks}\\
\hline 9.2 & \emph{books} , \emph{notebooks}\\
\hline ... & ...\\
\hline 9.30& pens\\
\hline ... & ...\\
\hline 2.1 & \emph{books} , \emph{notebooks}\\
\hline 2.2 & \emph{books} , \emph{notebooks}\\
\hline
\end{tabular}
\caption{Transactional database $DB$ in a bookstore}
\label{bookstore}

\end{center}
\vspace{-0.9cm}
\end{table}
%\vspace{-1cm}

%\vspace{-0.6cm}

We present the basic concepts related to cyclic association rules that will be of use in the remainder.
%\vspace{-0.6cm}
\subsubsection*{\textbf{Time unit }}
%\vspace{-0.35cm}
Considering the temporal aspect, the first considered measure is the time unit. Firstly, it was introduced by Odzen \emph{et al} \cite{CAR98}.

%\vspace{-0.2cm}
\begin{definition}
\par Given a transactional database $DB$, each time unit u$_{i}$
corresponds to the time scale on the database \cite{CAR98}.
\end{definition}
%\vspace{-0.6cm}
\begin{Example}
\par Let the following example be highlighted in table \ref{context1}.
%\vspace{-0.5cm}
\par \begin{table}[htpd]
%\vspace{-0.4cm}
\begin{center}
\begin{tabular}{| c | c |}
\hline \textbf{Transaction ID} & \textbf{Items}\\
\hline \hline 1 & B\\
\hline 2 & A, B \\
\hline 3 & A, B, C, D \\
\hline 4 & A, B, C \\
\hline 5 & C \\
\hline 6 & A \\
\hline
\end{tabular}
\caption{Initial database $DB$. \label{context1}}
%\vspace{-0.9cm}
\end{center}
\end{table}
%\vspace{-0.4cm}

\par The transactions illustrated in table \ref{context1} are extracted \emph{hourly}. So that, the corresponding time unit is the \emph{hour}.
\end{Example}
%\vspace{-0.7cm}
\subsubsection*{\textbf{Cycle}}
%\vspace{-0.1cm}

The concept of cycle was primarily introduced by Odzen \emph{et al} \cite{CAR98}.
%\vspace{-0.2cm}
\begin{definition}
\par A cycle c is a tuple $(l, o)$, such that $l$ is the length cycle,
being multiples of the unit of time; $o$ is an offset
designating the first time unit where the cycle appeared.\\
Thus, we conclude that 0 $\le$ $o$ $<$ $l$.
\end{definition}
%\vspace{-0.5cm}
\begin{Example}
\par If we consider a length of cycle $l$= 2 and the
corresponding offset is 1. So that, the cycle $c$ =($l, o$)=(2,1).
\end{Example}

%\vspace{-0.3cm}
Approaches addressing the issue of cyclic association rules are the following:
%\vspace{-0.3cm}
\begin{itemize}
%\vspace{-0.4cm}
\item The \textsc{\textbf{Sequential Approach}}: is a two-phase based algorithm. The key idea is: (\emph{i}) to generate the large itemsets and
to extract straightforwardly the corresponding association rules. (\emph{ii}) to detect the cycles of rules. So those are cyclic will be kept and the remainder is pruned.

\item The \textsc{\textbf{Interleaved Approach}}: is a three-phase based algorithm. Thanks to cycle pruning \footnote{Cycle Pruning is a technique for approximating the cycles of itemsets. In fact, we considerer : "If an itemset X has a cycle, then any of the subsets of X has the same cycle"}, we generate the potential cycles.
For every unit of time, we apply the cycle skipping \footnote{Cycle Skipping is a technique for avoiding
counting the support of an itemset in time units, which we
know,  cannot be part of a cycle of the itemset.} to extract the itemsets and we count their support thanks to the cycle elimination \footnote{Cycle Elimination is a technique relying on this property : "If the support for an itemset X is below the minimum
support threshold in time segment then X cannot have any of the cycles in sub time segments}

\item The \textsc{\textbf{PCAR Approach}}: Radically, it is based on the segmentation of the database in a number of partition fixed by the user.
The browse of the database is done sequentially partition by partition. This latter is scanned to generate the frequent cyclic itemsets.
%Obviously, the given partition $i$ will be scanned only once, the next partition $i+1$ will use the frequent cyclic itemsets already generated in the partition $i$ hence we emphasize the incremental side of our model.
Achieving the last one, we obtain the set of frequent cyclic itemsets. Hence, we extract from them the cyclic association rules.

%\vspace{-0.4cm}

\end{itemize}
%\vspace{-0.4cm}
\subsection{\textbf{Incremental mining of association rules}}

%\vspace{-0.3cm}

In the incremental context, several streams of approaches were reported to update incrementally the discovered association rules.
We start by introducing the most well-known ones.
%\item {\textbf{FUP}}
%\textsc{FUP} is based on the scan of the incremental database to generate frequent itemsets. If they weren't already frequent in the initial database, %a rescan of the initial database is crucial to decide if they can be considered as frequent itemsets or not according to a given threshold.

Initially, the key idea of incremental update was proposed by Cheung et al by introducing \textsc{FUP} for incrementally updating frequent itemsets \cite{Che96}. This approach assumes batch updates and
takes advantage of the relationship between the original database $DB$ and
the incrementally added transactions $db$.
Inspired from the \textsc{Apriori} algorithm, the key idea of \textsc{FUP} is that by adding db to $DB$, some previously frequent itemsets will remain frequent and some previously infrequent itemsets will
become frequent (these itemsets are called winners). At the same time, some
previously frequent itemsets will become infrequent (these itemsets are called
losers). The main contribution of \textsc{FUP} is to use information in $db$ to filter out
some winners and losers, and therefore reduce the size of candidate set in
the Apriori algorithm. Because the performance of the Apriori algorithm
relies heavily on the size of candidate set, \textsc{FUP} improves the performance of
\textsc{Apriori} greatly.

%Although the huge speed marking the \textsc{FUP} algorithm comparing to the Apriori, it suffers from some deficiencies.
%In fact, it was built on the Apriori algorithm with small modifications. So many passes of checking against the original database would usually be required.

The \textsc{BORDERS} Algorithm developed by Thomas \emph{et al.} \cite{26} and Feldman \emph{et al.}\cite{13}, is another approach using the concept of "\emph{negative border}" introduced by Toivonen \cite{27} aiming to indicate if it is necessary or not to check any candidate against the initial database. The main idea outlined by this algorithm is its need at most of one scan of the original database in the update operation. In this respect, the algorithm maintains information about the support of frequent itemsets in the original database along with the support of their negative border. If any itemset becomes a winner in the updated database, it follows that some itemset formerly in the negative border will also become a winner. Consequently, the negative border can be considered as an indicator for the necessity of looking for winners in the original database. If no expansion happens in the border, no need brings a point up to scan the original database.

%\par Therefore, the focus is put on the main advantage of this algorithm which is avoiding rescan of the original database allowing efficiently fast update operations.

\par Another strategy for maintaining association rules in dynamic databases, the {\textsc{Weight}} approach, is proposed by Shichao Zhang \emph{et al.}  \cite{Zha03}. This method used weighting technique to highlight new data. Indeed, it is a four-phase based approach: (i)Firstly, all frequent and
    hopeful itemsets related to the initial database are stored; (ii) Secondly, each incremental dataset DB$_{i}$ is mined and all frequent itemsets are stored. According to the requirements given by users \footnote{For example, a new or infrequent item can be expected to be completed as a
frequent item when it is strongly supported n times continually by incremental
data sets. This is an example of an outlined requirement}, an assignment of a weight to each set
    DB$_{i}$ is subjectively done; (iii) Thirdly, the aggregation of all rules based on hopeful itemsets by weighting is accomplished; (iv) Finally, a selection of high rank itemsets is achieved being the founded output.

%Nevertheless, the price of this privilege is paid during the original extraction because the initial extraction must be made according to a threshold of support less than the introduced parameter of support threshold. \\

 %\par Although the panoply of algorithms dedicated to the incremental maintenance issue of association rules, we assume there is no approach proposed in the scope of cyclic association rules.
% In this respect, we present a description of the incremental problem of cyclic association rules.

%In this paper, we focus on incremental maintenance of cyclic association. Indeed, a presentation of cyclic association rules seems fundamental being the main moan of the following subsection.

%\section{Foundations of the proposed algorithm}

%To sum up, (\emph{i}) the \textsc{Sequential} algorithm point out the independency between association rule mining and cycle detection, and (\emph{ii}) the \textsc{Interleaved} algorithm, employs optimization techniques on discovering cyclic association rules. As far the \textsc{PCAR} Algorithm is the fastest and amply outperforms the others.

%In this paper, we are interested in presenting a new model to incrementally update the cyclic association rules.
%First, we discuss the surveyed approaches dedicated to incremental maintenance. Then, we propose an algorithm to efficiently maintain the cyclic association rules.
%Through extensive carried out experiments on benchmarks and datasets, we stress on the effectiveness of our proposal on runtime performances.

%\section{\textbf{Related work}}

%\vspace{-0.5cm}

 \section{\textbf{Incremental mining of cyclic association rules}}

%\vspace{-0.42cm}

 Along this section, we present a description of the tackled problem. First, we formally define the problem of cyclic association rules. Then, we present the basic notions.
 %\vspace{-0.65cm}
\subsection{\textbf{Formal problem description}}
%\vspace{-0.3cm}
\par Regarding cyclic association rules, we stress on rules that are repeated in a cyclical way.
 Indeed, given a length of cycle, we extract itemsets that appear sequentially in the database. Let consider X and Y two itemsets appearing in DB at transaction number {i} and sequentially at the transaction number i+\textsc{length of cycle} until the end of the database(Table \ref{c1}). According to given support threshold, we prune the non frequent cyclic itemsets. Thus, we generate the cyclic association rules based on minimum confidence threshold.
%\vspace{-0.6cm}

\par \begin{table}[htpd]
\begin{center}
\begin{tabular}{| c | c |}
\hline \textbf{Transaction ID} & \textbf{Items}\\
\hline \hline i & X , Y\\
\hline ... & ...\\
\hline i+length of cycle & X , Y\\
\hline ... & ...\\
\hline i+(length of cycle*2) & X , Y\\
\hline ... & ...\\
\hline i+(length of cycle*k) & X , Y\\
\hline
\end{tabular}
\caption{cyclic itemsets in DB}
\label{c1}
\end{center}
%\vspace{-0.8cm}
\end{table}
%\vspace{-0.5cm}

To summarize, the cyclic association rules mining problem can be reduced to extraction of frequent cyclic itemsets, because once we have frequent cyclic itemsets set, cyclic association rules generation will be straightforward.
\par After several updates of $DB$, an increment $db$ of $\mid$db$\mid$ transactions is added to $DB$. The problem of incremental maintenance of cyclic association rules is to compute the new set of the frequent cyclic itemsets in $DB$'=$DB \cup db$ according to a support threshold $MinSup$.
\par In order to extract cyclic association rules from databases, the only plausible solution is to rerun one of the classical algorithms dedicated to the generation of cyclic association rules \emph{i.e.}, \textsc{Sequential}, \textsc{Interleaved} or \textsc{PCAR}. As a result, two drawbacks are quoted:
\begin{itemize}
%\vspace{-0.2cm}
\item If the original database is large, much computation time is wasted in maintaining association rules whenever new transactions are generated;
\item Information previously mined from the original database, provides no help in the maintenance process.
%\vspace{-0.3cm}
\end{itemize}

\vspace{-0.5cm}
\subsection{\textbf{Basic notions}}
\vspace{-0.3cm}
We start this subsection by presenting the key settings that will be of use in the remainder.
\vspace{-0.6cm}
\subsubsection*{\textbf{\emph{Frequent Cyclic itemset}}}
This concept refers to cyclic itemsets having supports exceeding the considered threshold. The formal definition is as follows.
\vspace{-0.6cm}
\par \begin{definition}
\par Let $XY$ be an itemset, the $sup(XY)$ is the support of the itemset in the database, reminding that only cyclic occurrences are considered on the support computing, and the minimum support threshold reminding $MinSup$.
The itemset $XY$ is considered as \emph{\textbf{Frequent Cyclic}}
denoted $\emph{\textbf{FC}}$ if the cyclic occurrences of the itemset
$XY$ are greater or equal to the given support threshold
otherwise if $sup(XY)$ $\geq$ $MinSup$.
\end{definition}
\vspace{-0.4cm}
\begin{Example}
\par We consider the context shown by table \ref{context1}, the $Minsup$ equal to 2 and the length of cycle is 2. The
binary sequence representing the itemset $AB$ is 011100 so $sup(AB)$=$MinSup$=2 then $AB$ is called \emph{\textbf{Frequent Cyclic itemset $FC$}}.
\end{Example}

%\vspace{-0.6cm}

\subsubsection*{\textbf{\emph{Frequent Pseudo-Cyclic itemset}}}
\vspace{-0.2cm}
The {\emph{frequent pseudo-cyclic}} concept is presented as follows.
\begin{definition}
\par Let $XY$ be an itemset, the $sup(XY)$ is the support of the itemset in the database, the $MinSup$ the minimum support threshold. The itemset $XY$ is considered as \textbf{\emph{frequent pseudo-cyclic}} denoted $\textbf{\emph{FPC}}$ if its support is less than $MinSup$.
Simultaneously, its support is greater than a given threshold called $\emph{\textbf{MinFPC}}$.
\[ MinFPC  \le  sup(XY) < MinSup  \]

\end{definition}
\vspace{-0.4cm}
\begin{Example}
\par Given the previous context, we consider $MinSup$ equal to 2, the $MinFPC$ is 0.2 and the
length of cycle is 2. The binary sequence representing the
itemset $AD$ is 000100 so $sup(AD)$ = 1 $<$ $MinSup$=2 $\geq$ $MinFPC$=0.2 then $AD$ is called \textbf{\emph{Frequent Pseudo-Cyclic itemset }}$\textbf{\emph{FPC}}$.
\end{Example}

%\vspace{-0.6cm}
\subsubsection*{\textbf{\emph{Minimum $\textbf{FPC}$ threshold}}}
According to this measure, we classify the remainder of the itemsets after $MinSup$ pruning on hopeful cyclic itemsets that are not frequent in the initial database but are more likely to move to this status in the increment database.
\begin{definition}
The \emph{\textbf{Minimum $\textbf{FPC}$} \textbf{threshold}}, denoted by $\emph{\textbf{MinFPC}}$, refers to a threshold dedicated to prune the none hopeful itemsets. It is computed according to this formula:
\vspace{-0.6cm}
\par
\[  MinFPC=
\frac{\frac{MinSup}{\mid{DB}\mid+\mid{db}\mid}+MinSup}{\mid{DB}\mid + \mid{db}\mid }
\]
\vspace{-0.6cm}
\end{definition}
\vspace{-0.45cm}
\begin{Example}
\par Continuing with the same database $DB$ considered as initial one, let the database $db$ containing 4 transactions be the increment one. In addition we fix the $MinSup$ to 2. Then the \emph{\textbf{$MinFPC$}} is
computed as follows:
\vspace{-0.35cm}
\[  MinFPC=
\frac{\frac{2}{\mid{6}\mid+\mid{4}\mid}+2}{\mid{6}\mid + \mid{4}\mid}=0.2
\]
\vspace{-0.6cm}
\end{Example}
%\vspace{-0.6cm}

\subsubsection*{\textbf{\emph{Non Frequent Cyclic Itemset}}}
\vspace{-0.3cm}
This concept refers to cyclic itemsets that are not both frequent cyclic itemsets and frequent pseudo-cyclic itemsets.
%\vspace{-0.28cm}
\begin{definition}
\par Let $XY$ be an itemset, the $sup(XY)$ is the support of the itemset in the database and the $MinSup$ the minimum support threshold.
The itemset $XY$ is considered as \textbf{\emph{non frequent cyclic itemset} }denoted $\textbf{\emph{NFC}}$ if the support of of $XY$ is less than the given $MinFPC$ threshold
otherwise if $sup(XY)$ $<$ $MinFPC$.
\end{definition}
\vspace{-0.4cm}
\begin{Example}
\par Given the previous context, we consider $MinSup$ equal
to 4, the $\emph{\textbf{MinFPC}}$ is 2 and the length of cycle is 2. The
binary sequence representing the itemset $AD$ is 000010 so
$sup(AD)$=1 $<$ $MinFPC$=2 $<$ $MinSup$=4 then $AD$ is called \textbf{\emph{non frequent cyclic itemset}} denoted $\textbf{\emph{NFC}}$.
\end{Example}

%\vspace{-0.4cm}

In this respect, the main thrust of this paper is to propose a new strategy dedicated to the incremental update of cyclic association rules aiming to reduce efficiently the runtime required for the generation of cyclic association rules in the case of addition of transactions at the maintenance process of databases. Indeed, this proposal is outlined in the following section.

%\vspace{-0.6cm}
\section{\textbf{IUPCAR Algorithm}}
%\vspace{-0.4cm}
In order to maintain incrementally the cyclic association rules, we introduce a novel approach called \textsc{Incremental UPdate of Cyclic Association Rules} denoted \textsc{IUPCAR}. Indeed, the \textsc{IUPCAR} algorithm operates in three phases:
 \begin{itemize}
 %\vspace{-0.3cm}
\item In the first phase, a scan of the initial database is done to class the founded itemsets on three classes namely the frequent cyclic itemsets, the frequent pseudo-cyclic itemsets and non frequent cyclic itemsets.
\item In the second phase, according to the second database, we categorize the itemsets into the three mentioned classes.
Then, depending of the ancient class of the itemset with its ancient support and the new class with its new support in the increment database,
 an affectation of the suitable class is made according to a weighting model.
\item In the final phase, given the founded frequent cyclic itemsets after the update operation, the corresponding cyclic association rules are generated.
%\vspace{-0.3cm}
\end{itemize}
%%\vspace{-0.7cm}

\begin{figure}[htpd]
    \centering
\begin{tabular}{c}
\includegraphics[width=3in, height = 1.75in]{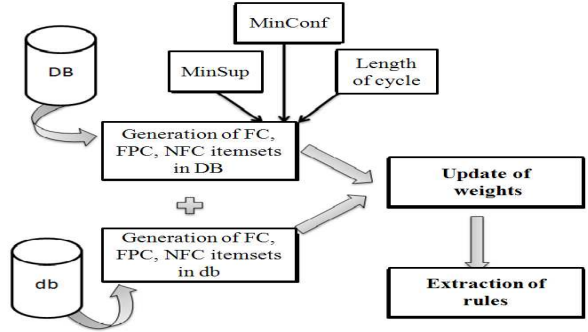} \\  % should insert INTERVAL AR not R1
%\hline
\end{tabular}
\caption{ The flowchart of \textsc{IUPCAR}.} \label{process_iupcar}
\end{figure}

%%\vspace{-0.2cm}

As highlighted by figure \ref{process_iupcar}, first and foremost, the \textsc{IUPCAR} algorithm takes on input the initial database, the minimum support threshold $MinSup$, the minimum confidence threshold $MinConf$ and the length of cycle. According to those key settings, a generation of frequent cyclic itemsets, frequent pseudo-cyclic itemsets and non frequent cyclic itemsets from the initial transactions is done.
Stressing on the dynamistic feature of the databases, we add the novel transactions building the $db$ database. To accomplish this update operation, a scan of the new database is done and a generation of the itemsets and their classification are straightforwardly realized. After that, an update of the status and the weights of itemsets are done without rescanning the initial database. Finally, we generate the cyclic association rules based on the retained frequent cyclic itemsets.

\par  Intuitively in the updating problem, we assume the following cases shown by table \ref{cases}:
\begin{itemize}
%\vspace{-0.2cm}
\item Frequent cyclic itemset $FC$ is already saved as frequent cyclic $FC$ (case \textbf{A}), frequent pseudo-cyclic $FPC$ (case \textbf{B}) or non frequent cyclic itemset $CNF$ (case \textbf{C});
\item Frequent pseudo-cyclic itemset $FPC$ is already saved as a frequent cyclic $FC$(case \textbf{D}), frequent pseudo-cyclic $FPC$ (case \textbf{E}) or non frequent cyclic itemset $NFC$ (case \textbf{F});
\item Non frequent cyclic itemset $NFC$ is already saved as frequent cyclic $FC$(case \textbf{G}), frequent pseudo-cyclic $FPC$ (case \textbf{H}) or non frequent cyclic itemset $NFC$ (case \textbf{J}).
    %\vspace{-0.4cm}
\end{itemize}
%\vspace{-0.5cm}

\begin{table}[htpd]
\begin{center}
{\small
\begin{tabular}{| c | c | c | c |}
\hline $db$-$DB$ & $FC$  & $FPC$ & $NFC$\\
\hline $FC$ & Always $FC$  & Computation based& Computation based\\
&& on db and DB &  on db and DB  \\
& (case \textbf{A}) & (case \textbf{B}) &  (case \textbf{C}) \\
\hline $FPC$ & Computation based& Always $FPC$  & Computation based\\
& on db and DB &&on db and DB  \\
& (case \textbf{D}) & (case \textbf{E}) &  (case \textbf{F}) \\
\hline $NFC$ & Computation based& Computation based& Always $NFC$  \\
&on db and DB &on db and DB&  \\
& (case \textbf{G}) & (case \textbf{H}) &  (case \textbf{J}) \\
\hline
\end{tabular}

\caption{Possible cases in update operation.}\label{cases}
}
\vspace{-0.7cm}

\end{center}
\end{table}
%%\vspace{-0.1cm}

%\vspace{-0.8cm}

\par To handle those various cases, we introduce the following weighting model.

%\subsubsection{\textbf{Weighting model}}
In the update operation, a dramatic change in the status of the itemsets between the first and the coming database is intuitively plausible. That's why, we refer to the weighting model as a technique dedicated to decide which status is the suitable to the itemset after the new added transactions. Indeed, we sketch the mechanism of weighting model as follows.
%\newpage
%\vspace{-1.5cm}

For an itemset $X$, we :
 \begin{itemize}
\vspace{-0.2cm}

\item \emph{compute} the relative support of $X$ in the initial database $DB$, denoted by $Sup(X_{DB}$), according to to the given formula:
\vspace{-0.3cm}

\[Sup(X_{DB})=\frac{Sup(X)}{|DB|}\]
\vspace{-0.5cm}

\item \emph{compute} the relative support of $X$ in the increment database $db$, denoted by $Sup(X_{db}$), according to the given formula:
\vspace{-0.3cm}

\[Sup(X_{db})=\frac{Sup(X)}{|db|}\]
\vspace{-0.5cm}

\item \emph{compare} the relative support of $X$ in the initial database $Sup(X_{DB})$ \emph{vs.} that of the increment database $Sup(X_{db})$. And we \emph{choose the greatest one}.
\item Two alternatives are plausible :
\begin{enumerate}
\item If the itemset has the same state in the initial and the incremental database,
 we will \textbf{enhance} its weight;
\item If the state of the itemset has changed from the initial to the incremental database,
 we will \textbf{check} which one of its states has the greatest weight and we will \textbf{decrease} its value and \textbf{affect} this state as its new one.\\
\end{enumerate}
%\vspace{-0.4cm}

 In this respect, let the new weight of $X$ be denoted by \textbf{$\mathcal W(_{X_{db} })$}.\\
 The table \ref{cases} sketches the possible cases that can be summarized on three possible scenarii:

 \begin{enumerate}
 \item No change in the status simply happens. So, the itemset remains frequent cyclic $FC$ (case \textbf{A}) or frequent pseudo-cyclic itemset $FPC$ (case \textbf{E}) or non cyclic frequent $CNF$ (case \textbf{J}). The new weight is computed as follows:
     \vspace{-0.6cm}

 \[ \mathcal W(_{X_{db} }) = \frac{\mathcal Sup(X_{DB} ) + \mathcal Sup(X_{db} ) }{\mid {\mathcal DB}\mid +\mid {db}\mid } \]
%\vspace{-0.5cm}

\item A change in the status between the initial transactions and the new ones occurs. So one of the cases depicted on the table \ref{cases} by (case \textbf{B}), (case \textbf{C}), (case \textbf{D}), (case \textbf{F}), (case \textbf{G}) or (case \textbf{H}) happens. Then, two situations are obviously outlined:
     \begin{enumerate}
     \item    If \emph{\textbf{the previous support of the
         itemset is greater than the new one in the
         increment database}}, the affected status is remained the same and its novel weight is computed as follows:
\vspace{-0.4cm}

\[\mathcal W(X_{db}) =
\frac{\mathcal Sup(X_{DB})}{\mid {\mathcal DB}\mid } - \frac {\mathcal Sup(X_{db}) }{\mid {db}\mid}
\]
\vspace{-0.4cm}

     \item If \emph{\textbf{the new support of the itemset
         is greater than the previous one}}, the affected status is the new one and its novel weight is computed as follows:
        \vspace{-0.4cm}

\[\mathcal W(X_{db}) =\frac{\mathcal Sup(X_{db})} {\mid {db}\mid }- \frac{\mathcal Sup(X_{DB}) }{\mid
{\mathcal DB}\mid }\]
\vspace{-0.5cm}

 \end{enumerate}
\end{enumerate}
\item Considering the update operation of the itemsets' status and weights, we extract cyclic association rules based on frequent cyclic itemsets.

\end{itemize}

%\vspace{-0.8cm}

\section{\textbf{IUPCAR Example}}
%\vspace{-0.3cm}
\par  Aiming to illustrate deeply the mechanism of our approach with its different steps,
we consider the context sketched in table \ref{context} as an initial database.
%%\vspace{-0.7cm}
%%\vspace{-0.8cm}

%\renewcommand{\labelitemi}{$\triangleright$}
\par We introduce the following parameters:
%\vspace{-0.2cm}

\begin{itemize}
\item	Length of cycle equal to 2;
\item	$MinSup$ equal to 50\%;
\item $\mid$$db$$\mid$ equal to 4.
\end{itemize}
%\vspace{-0.4cm}

%\renewcommand{\labelitemi}{$\bullet$}
\par We propose to illustrate the possible cases, we choose one itemset to facilitate the explanation of our proposal.
%\newpage
Indeed, based on the initial database, we can extract :
\begin{itemize}

%\vspace{-0.2cm}
\item classified as $FC$: AB;
\item classified as $FPC$: AC;
\item classified as $NFC$: AD.
%\vspace{-0.4cm}

\end{itemize}
%\vspace{-0.4cm}
\par \begin{table}[htpd]
\begin{center}
\begin{tabular}{| c | c |}
\hline \textbf{Transaction ID} & \textbf{Items}\\
\hline \hline 1 & B\\
\hline 2 & A, B \\
\hline 3 & A, B, C, D \\
\hline 4 & A, B, C \\
\hline 5 & C \\
\hline 6 & A \\
\hline
\end{tabular}
\caption{Initial database $DB$. \label{context}}
%\vspace{-0.6cm}

\end{center}
\end{table}
%\vspace{-0.8cm}

\par For the itemset AB, recognized as $FC$, we simulate the various cases that can be handled in the update operation. Furthermore, the table \ref{tab_AB} summarized the possible new status of the itemset AB and its eventual supports in $db$.
%\newpage
\par According to $db$, we find :
\begin{enumerate}
%\vspace{-0.2cm}

\item \textbf{\emph{Scenario (\textbf{a})}}: \emph{AB is generated as a $FC$}: \\
W($_{AB_{db} }$) =  $\frac{Sup(AB_{DB}) + Sup(AB_{db}) }{|DB|+|db|}$
    = $\frac{3 + 2 }{6 + 4}$ =$\frac{1}{2}$.\\
    So that the new state affected is clearly $FC$ but the weight of AB is increased due to its keeping the same status in $DB$ and $db$;

\item \textbf{\emph{Scenarii (\textbf{b, c})}}: \emph{AB is generated as a $FPC$}:\\
    \begin{enumerate}
    \item \textbf{\emph{Scenario (\textbf{b})}}: the support of AB in $DB$ is greater than the support of AB in $db$ \\
    W($_{AB_{db} }$) = $\frac{Sup(_{AB_{DB} })}{|DB|}$ - $\frac{Sup(_{AB_{db} }) }{|db|}$ = $\frac{1}{2}$-$\frac{1}{4}$=$\frac{1}{4}$;

    \item \textbf{\emph{Scenario (\textbf{c})}}: the support of AB in $DB$ is less than the support of AB in $db$\\
    W($_{AB_{db} }$) = $\frac{Sup(_{AB_{db} })}{|db|}$ - $\frac{Sup(_{AB_{DB} }) }{|DB|}$ = $\frac{3}{4}$-$\frac{1}{2}$=$\frac{1}{4}$.

\end{enumerate}
%%\vspace{-2m}
%\vspace{-0.7cm}

\par \begin{table}[htpd]
\begin{center}
\begin{tabular}{| c | c | c | c | c | c |}
\hline \textbf{AB} & \textbf{$FC$}& \multicolumn{2}{|c|} {\textbf{$FPC$}}& \multicolumn{2}{|c|}{ \textbf{$NFC$}}\\
\hline \hline
$Sup(AB_{db})$ &2 &1 &3 &1&3 \\
&Scenario \textbf{a}&Scenario \textbf{b}&Scenario \textbf{c}&Scenario \textbf{d}&Scenario \textbf{e}\\
\hline
\end{tabular}
\caption{The possible cases in $db$ for $FC$ itemset. \label{tab_AB}}
\end{center}
%\vspace{-0.4cm}

\end{table}
%\vspace{-0.7cm}

\item \textbf{\emph{Scenarii (\textbf{d, e})}}: \emph{AB is generated as a $NFC$}:
\begin{enumerate}
    \item \textbf{\emph{Scenario (\textbf{d})}}: the support of AB in $DB$ is greater than the support of AB in $db$\\
    W($_{AB_{db} }$) = $\frac{Sup(_{AB_{DB} })}{|DB|}$ - $\frac{Sup(_{AB_{db} }) }{|db|}$ = $\frac{1}{2}$-$\frac{1}{4}$=$\frac{1}{4}$;
    \item \textbf{\emph{Scenario (\textbf{e})}}: the support of AB in $DB$ is less than the support of AB in $db$\\
    W($_{AB_{db} }$) = $\frac{Sup(_{AB_{db} })}{|db|}$ - $\frac{Sup(_{AB_{DB} }) }{|DB|}$ = $\frac{3}{4}$-$\frac{1}{2}$=$\frac{1}{4}$.

\end{enumerate}
\end{enumerate}
%\vspace{-0.2cm}

\par Consequently, the new state affected to $AB$ is $FC$ ({Scenarii (\textbf{b, d})}) because its support in $DB$ is greater than its one in $db$. Nevertheless, the new affected status will have a weight less than the previous support because in the incremental database, we notice the change of its status;
\par Likewise, we affect for $AB$ $FPC$ or $FCN$ ({Scenarii (\textbf{c, e})}) as new status because its support in the incremental database is greater than its one in the initial database. However, the new affected status will have a weight less than the one extracted in the incremental database because it does not maintain its initial status.

\par Similarly, the new two states of itemsets $AC$ and $AD$ are respectively as $FPC$ and $NFC$, the same possible scenarii simulated for $AB$ are obviously in the update operation plausible.

\par As a final step, after the update of the status of the itemsets and their weights, only frequent cyclic itemsets are considered in the extraction of the novel cyclic association rules related to the both databases $DB$ and $db$.

%\vspace{-0.6cm}

\section{\textbf{Experimental study}}
%\vspace{-0.4cm}

\begin{figure}[htpd]
    \centering
\begin{tabular}{ c}
\includegraphics[width=2.5in, height = 2in]{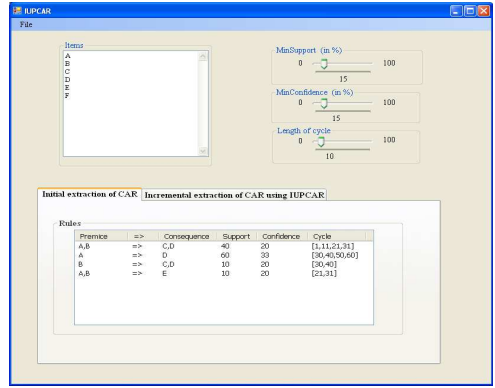} \\  % should insert INTERVAL AR not R1

\end{tabular}
%\vspace{-0.3cm}

\caption{ The user interface design of \textsc{IUPCAR}.} \label{up}
%\vspace{-0.7cm}

\end{figure}

To assess the IUPCAR efficiency, we conducted several experiments on a PC equipped with a 3GHz Pentium IV and 2GB of main memory.
The figure \ref{up} is an illustration of the \textsc{IUPCAR} user interface.

During the carried out experimentation, we used benchmarks datasets taken from the UC
Irvine Machine Learning Database Repository.
%%\vspace{-0.5cm}

\begin{table}[ht]
 \begin{center}
 %  \tabcolsep = 2\tabcolsep
   \begin{tabular}{|l|c|c|c|c|}
\hline \hline
\textbf{Database} & \textbf{$\sharp$Transactions} & \textbf{$\sharp$Items} & \textbf{Average size} & \textbf{Size(Ko)}  \\
& & & \textbf{of transactions} & \\
\hline \hline
 \textsc{T10I4D100K} & 100000 & 1000 & 10 & 3830 \\
 \textsc{T40I10D100K} & 100000 & 775 & 40 & 15038 \\
 \textsc{Retail} & 88162 & 16470 & 10 & 4070 \\
\hline
\end{tabular}
\caption{Description of benchmark
databases.\label{tab_Characteristics}}
\end{center}
\end{table}
%%\vspace{-1cm}

Table \ref{tab_Characteristics} depicts the characteristics of the datasets used in our evaluation.
It shows the number of items, the number of transactions, the average size of transactions and the size of each database.

Through these experiments, we have a twofold aim: first, we have to stress on the performance of our proposal by the variation of $MinSup$ on the one hand and the variation of the cardinalities of initial and incremental databases on the other hand. Second, we put the focus on the efficiency of our approach {\emph{vs.}} that proposed by the related approaches of the literature.
%\vspace{-0.7cm}

%\vspace{-0.5cm}

\subsection{\textbf{Performance aspect}}
%\vspace{-0.4cm}

%\begin{itemize}
%\item
{\textbf{Minimum support variation}}
%%\vspace{-0.1cm}

In the carried out experimentations, we divided database in two partitions: $DB$ is the initial database and $db$ is the incremental one.
Firstly, the $DB$ is constituted of 70\% of the size of the benchmark dataset and $db$ is constituted of the remainder namely the 30\%.
Secondly, we increase the size of the initial database to achieve 80\% from the size of the benchmark dataset so the $db$ presents only 20\%.
To finish with an initial database representing 90\% and an incremental one providing only 10\%.
%%\vspace{-0.7cm}

\begin{figure}[htbp]
    \centering
\begin{tabular}{c c c}
\includegraphics[width=1.5in, height = 1.5in]{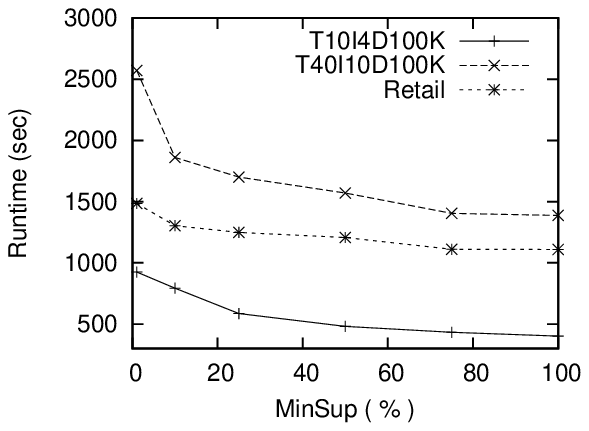} &
\includegraphics[width=1.5in, height = 1.5in]{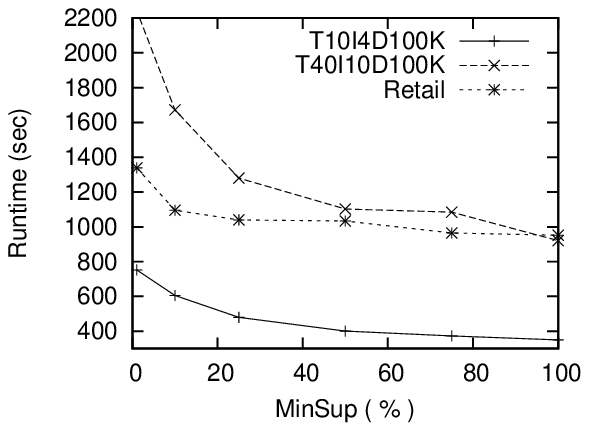} &
\includegraphics[width=1.5in, height = 1.5in]{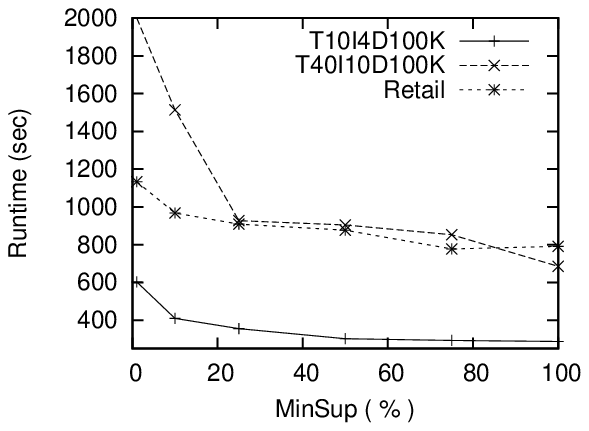} \\

($DB$=70\%, $db$=30\%) & ($DB$=80\%, $db$=20\%) &($DB$=90\%, $db$=10\%)\\
\end{tabular}
\caption{Experimental results of \textsc{IUPCAR} for an incremental database =10\%, 20\% or 30\%.}
\label{xp20}
\end{figure}
%%\vspace{-0.9cm}

%\paragraph*{$\bullet$ DB=70\% and db=30\%}
\par Considering the given parameters : the $MinConf$ =50\%,
 the length of cycle =30, we present the variation of $MinSup$ and the corresponding
 runtime of \textsc{IUPCAR} in figure \ref{xp20}.

\par Indeed, by varying the support, it is obvious that the more support is increasing the more runtime of \textsc{IUPCAR} decreases.
For the \textsc{T10I4D100K} dataset, having $DB$=70\% and $db$=30\%, the runtime of \textsc{IUPCAR}
increases from 926,295 seconds for 1\% as a $MinSup$ to
482,726 seconds for 50\% as a $MinSup$, to stabilize at
around 400, for $MinSup$ exceeding 50\%.
As expected, this fact is similarly conceivable for \textsc{T40I10D100K} and \textsc{Retail} datasets.
 \par As shown figure \ref{xp20}, we assume worthily that on whatever the size of the initial and increment database,
 the more support increases the more runtime required for \textsc{IUPCAR} goes down.
{\textbf{Variation of the size of updated database}}

\par  In this part, we concentrate on the effect of variation of initial and incremental databases sizes.
%\begin{figure}[htbp]

\par
\begin{figure}[htbp]
   \centering
\begin{tabular}{ccc}
  \includegraphics[width=1.5in, height = 1.5in]{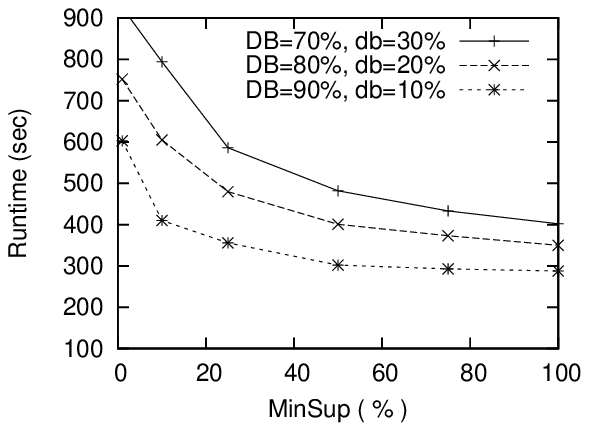}
&  \includegraphics[width=1.5in, height = 1.5in]{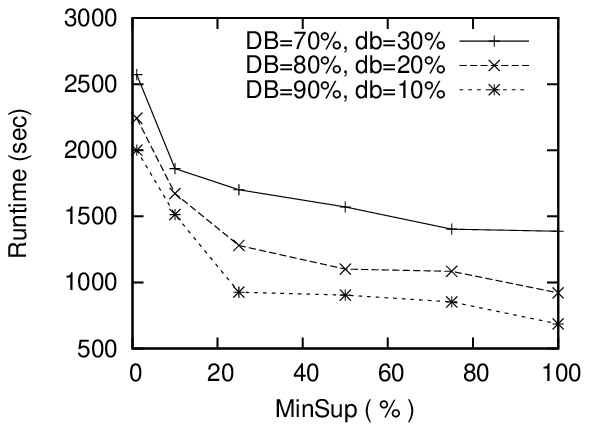}

& \includegraphics[width=1.5in, height = 1.5in]{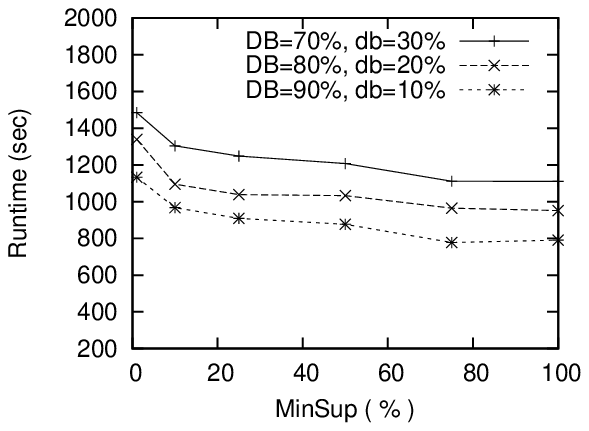} \\
(\textsc{T10I4D100K}) & (\textsc{T40I10D100K})  &(\textsc{Retail})\\
\end{tabular}
\caption{Comparing the runtime of \textsc{IUPCAR} with different incremental databases 10\%, 20\% and 30\% .}
%\vspace{-0.8cm}

\label{comparative}
\end{figure}

 Indeed, fixing parameters as follows: $MinConf$ equal
 to 50\% and length of cycle=3, according to the figure
 \ref{comparative}, we perceive regarding the \textsc{T40I10D100K} dataset for the same value of
 support equal to 50\%, with $DB$=70\% and $db$=30\% the update operation requires 1571,541 seconds but this value
goes down if we rise the size of the initial database
and we diminish the size of the increment one. Identically,
we notice 1102,84 seconds if the $DB$=80\% and $db$=20\% by
far 904,254 seconds if the $DB$=90\% and $db$=10\% of the
whole dataset.

\par  Therefore, it is crucial to deduce that the least is the size of the incremental database,
the least is the runtime required to update the cyclic
 rules and this can be noted for the two other datasets namely \textsc{retail} and \textsc{T10I4D100K}.
\par
%\vspace{-0.4cm}

%\end{itemize}
\subsection{\textbf{Efficiency aspect}}
%\vspace{-0.4cm}

\par  In order to evaluate the efficiency of our algorithm, we conducted comprehensive experiments to compare \textsc{IUPCAR} with the most efficient classical algorithm dedicated to cyclic association rules extraction namely \textsc{PCAR} algorithm.
The following values of parameters are set during the several experiences: the minimum of confidence equal to
50\%, the length of the cycle equal to 30 and the runtime of the algorithms regarding the \textsc{T10I4D100K}, \textsc{T40I10D100K} and \textsc{Retail} datasets.

% \subsection*{$\bullet$ DB=90\% and db=10\%}
 \par
\begin{figure}[htbp]
  \begin{center}
  \begin{tabular}{cc}
    \includegraphics[width=1.5in, height = 1.5in]{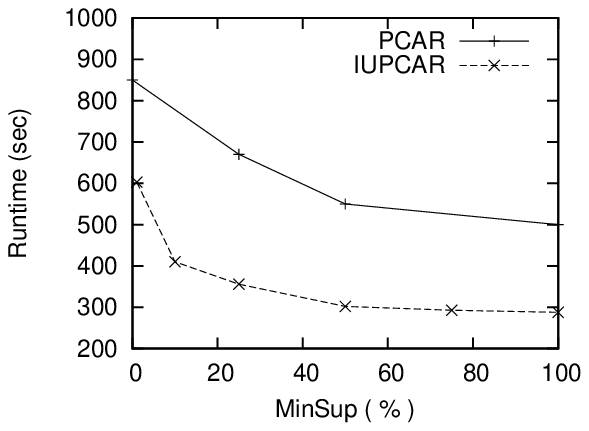}
  &  \includegraphics[width=1.5in, height = 1.5in]{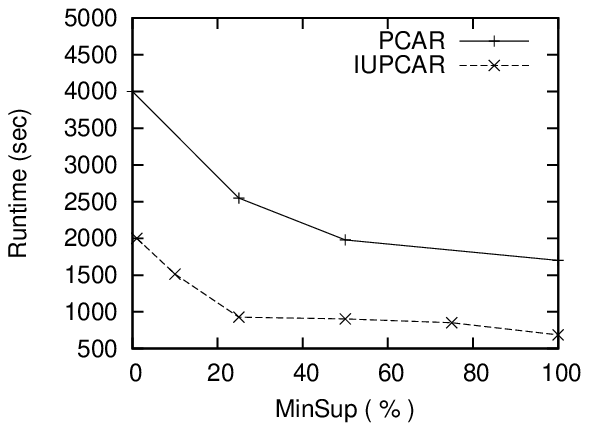} \\
  (\textsc{T10I4D100K}) & (\textsc{T40I10D100K})
  \end{tabular}
  \begin{tabular}{c}
   \includegraphics[width=1.5in, height = 1.5in]{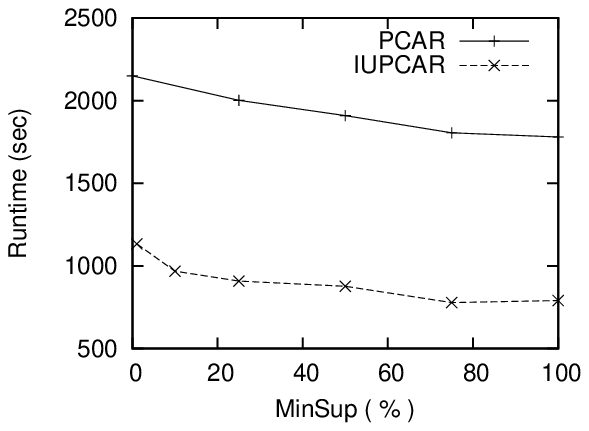} \\
   (\textsc{Retail})
  \end{tabular}
  \end{center}
  %\vspace{-0.6cm}

   \caption{Comparing the runtime of \textsc{IUPCAR} and \textsc{PCAR} with incremental database 10\%.}
     %\vspace{-0.6cm}

\label{fig10per} \end{figure}

  The results of varying the minimum support
  on running the \textsc{PCAR} algorithm and the
  \textsc{IUPCAR} one are shown by figure
  \ref{fig10per}.
  %\vspace{-0.8cm}

%%%%%%%%%%%%% there %%%%%%%%%%%%%%%%%%%%

%%\vspace{-0.5cm}

  \par  It indicates that the update operation of $DB$ by the
  increment $db$=10\% with a minimum support =100\%,
  requires 511,388 seconds by running \textsc{PCAR} and
  only its half 288,62 seconds by running \textsc{IUPCAR}
  for \textsc{T10I4D100K}dataset.

 \par For \textsc{T40I10D100K}, the original database $DB$ =90\% with the increment database $db$ =10\%
 required with a minimum support=100\% for
 \textsc{T40I10D100K} by running \textsc{PCAR} 1776,12
 seconds and interestingly its half 686,884 seconds for \textsc{IUPCAR} running
 .

 \par Similarly for \textsc{Retail}, the updating operation of the initial database
 by adding 10\% of its size with a minimum support=100\%
 requires 1894,416 seconds by running \textsc{PCAR} and efficiently
 791,9 seconds for \textsc{IUPCAR} running.

%refaire

%%\vspace{-0.4cm}

%\bigskip
\par Obviously, \textsc{IUPCAR} amply outperforms the \textsc{PCAR} algorithm in the context of maintenance of cyclic association rules and  proves its efficiency in various test cases.

%\vspace{-0.7cm}

\section{\textbf{Conclusion and perspectives}}
%\vspace{-0.4cm}

\par In this paper, we introduced the problem of incremental maintenance of cyclic association rules.
Thus, the flying over the pioneering approaches handling the incremental
update of association rules issue \cite{Lee97maintenanceof} conducted us to introduce a new proposal called \textsc{IUPCAR} algorithm dedicated particularly to update the cyclic association rules.
To evaluate its efficiency, several experimentations of the proposed method are carried out. So that, encouraging results are obtained.
Future work will focus mainly on : (\emph{i}) the quality of the generated cyclic association rules. In fact, we
plan to study deeply the significance of the extracted cyclic association rules for human experts \cite{Tse07} \cite{Yeh08},
(\emph{ii}) tackling the change support threshold in the incremental update operation of cyclic association rules \cite{Che99},
(\emph{iii}) using database vertical representation (Eclat (Zaki et al, 1997) \cite{zaki97}) to improve the \textsc{IUPCAR} results.
%\vspace{-0.6cm}

\end{document}